\documentclass[12pt]{article}

\usepackage{cite}
\usepackage{epsf}
\usepackage{graphicx}
\usepackage{psfrag}
\usepackage[dvips]{epsfig}

\usepackage[T2A]{fontenc}
\usepackage[cp866]{inputenc}
\usepackage[english]{babel}

\usepackage{amssymb,indentfirst}

\usepackage{amsmath}

\usepackage{verbatim}

\makeatletter
\renewcommand \thesection {\@arabic\c@section.}
\renewcommand\thesubsection   {\thesection\@arabic\c@subsection.}
\renewcommand\thesubsubsection{\thesubsection\@arabic\c@subsubsection.}
\makeatother

\def\starup#1{\mbox{$\raise1.8ex\hbox{$*$} \kern-.7em#1$}}
\def\krup#1{\mbox{$\raise1.8ex\hbox{$+$} \kern-1.0em#1$}}
\def\linup#1{\mbox{$\raise1.9ex\hbox{---} \kern-1.0em#1$}}

\begin{document}
\title{Colored scalar particles production in $pp$-collisions 
and possible mass limits for scalar gluons \\ 
 from future LHC data} 

\author{M.V.~Martynov\footnote{E-mail: martmix@mail.ru}, \,
 A.D.~Smirnov\footnote{E-mail: asmirnov@univ.uniyar.ac.ru}\\
{\small Division of Theoretical Physics, Department of Physics,}\\
{\small Yaroslavl State University, Sovietskaya 14,}\\
{\small 150000 Yaroslavl, Russia.}}
%\address{}%
%\email{}%
\date{}
\maketitle
%\thanks{}%

%\vspase
% ----------------------------------------------------------------

\begin{abstract}
\noindent
Cross sections of the colored scalar particle production in $pp$-collisions 
are calculated and differential and total cross sections of the 
corresponding parton subprocesses 
%$(gg\to\Phi\Phi^*)$ and $(q\bar{q}\to\Phi\Phi^*)$ 
are obtained. The total cross section of scalar gluon production 
in $pp$-collisions at the LHC is estimated and the dominant decays of scalar gluons are discussed. 
The production cross section of scalar gluons $F$ with masses 
$m_F\lesssim 1300$  GeV is shown to be sufficient for the effective  
production of these particles at the LHC.

\vspace{5mm}
\noindent
Keywords: Beyond the SM; four-color symmetry; Pati--Salam;
scalar leptoquarks; scalar gluons; scalar octets.

\noindent
PACS number: 12.60.-i

\end{abstract}

%\newpage

%-----------------------------------------------------------------

%\setlength{\baselineskip}{24pt}

The search for new physics beyond the Standard Model induced by higher 
symmetries (supersymmetry, left-right symmetry, etc.) is one of the modern 
research directions in elementary particle physics. One such symmetry 
possibly existing in nature is the four-color symmetry of quarks and 
leptons, where leptons are considered as the fourth color 
\cite{pati_salam}.

Under its own minimal unification with the symmetry of the Standard Model 
based on the gauge group 
\begin{equation}
SU_V(4)\!\times\! SU_L(2)\!\times\! U_R(1)
\end{equation}
(the minimal quark-lepton-symmetric model - MQLS-model \cite{smirnov-1995-346,smirnov-yadphys-1995}), 
the four-color symmetry under the Higgs mechanism of quark and lepton mass splitting predicts, 
in particular, new scalar particles, $SU_c(3)$ octets and $SU_L(2)$ doublets, 
the so-called scalar gluons $F_{a'i}$ ($a'=1,2$, $i=1..8$ are $SU_L(2)$ and 
$SU_c(3)$ indices respectively) \cite{smirnov-1995-346, smirnov-yadphys-1995, povarov-smirnov-yadphys-2001}.

By virtue of their Higgs origin, these particles have quark coupling 
constants proportional to the ratio of quark masses to vacuum expectation 
value of SM $\eta$ and, hence, the values of these constants are known 
(correct to within the quark mixing parameter), thus determining their 
interaction with gluons by the known strong coupling constant $g_{st}$. 
This makes it possible to quantitatively estimate the cross sections of 
processes in which these particles participate.
Estimates of contributions by the scalar gluon doublets $F_{a'i}$ 
to the Peskin-Takeuchi S-,T-,U-parameters of radiative corrections, and their comparisons 
with experimental data on S-,T-,U show that these scalar gluons may be relatively light 
(with masses of order 1 TeV or less) \cite{smirnov-2002-531,smirnov-yadphys-2003}.
Scalar gluons with such masses, being colored objects of the $SU_c(3)$ 
group, can be pair produced in $pp$-collisions through gluon fusion and, 
partly, through quark-antiquark pair annihilation. 
The scalar $SU_c(3)$-octets with natural suppression of FCNC 
%and their production at LHC 
%have been considered in Ref.~\cite{manohar-2006-74}. 
have been considered in Ref.~\cite{MW}. 
The production of scalar $SU_c(3)$-octets at LHC were considered in Refs.\cite{MW, GrWise, Gerbush, Zerwekh}. 
The light colored scalar octets in context of Adjoint $SU(5)$ unification have been discussed 
in Ref.~\cite{Perez}. 
In the case of the scalar gluons $F_{a'i}$ of MQLS-model the dominant decays of these particles  
are known \cite{popov-2005-20, PPSmPhAN2007}, which gives the possibility of search for 
these particles through their decays at LHC. 

%The dominant decays of the scalar gluon $F_{a'i}$ are known 
%\cite{popov-2005-20}, which gives the possibility of search for 
%these particles through their decays at LHC.

In the present paper we calculate the production cross sections of scalar particles 
of an arbitrary color representation in $pp$-collisions, 
and estimate these cross sections for $F_{a'i}$-scalar gluons at LHC energies 
with discussing the possible constraints on the masses of these particles 
from future LHC data with account of their dominant decay modes.

%The scalar $SU_c(3)$-octets with natural suppression of FCNC and their production 
%at LHC have been also considered in Ref.~\cite{manohar-2006-74}. 

%
%\vspace{10mm}
%\textbf{Production of colored scalar particles in $pp$-collisions}

Interactions of colored scalar particles $\Phi_i$ with gluons is contained 
in lagrangian
\begin{equation}
    \mathcal{L}_{\Phi\Phi g}
 = \sum_{scalars} \left [
\left (D^{\mu}_{ij} \Phi^j \right )^{\dagger}
                                 \left (D_{\mu}^{ik} \Phi_k \right )
 - M_\Phi^2 \Phi^{i \dagger} \Phi_i \right ],
\end{equation}

\begin{equation}
D_{\mu}^{ij}\Phi_j\ = \left(\partial_{\mu}  \delta^{ij} - i g_s
 {G}^a_{\mu}T_a^{ij}\right)\Phi_j,
\end{equation}
where $T_a^{ij}$ are the generators of the group representation $SU_c(N)$ 
($a$=$1,2..d_A$, $d_A$ - dimension of the adjoint representation of 
$SU_c(N)$), realized by the multiplets $\Phi_i$, $i,j$ - color index; in particular (for $N=3$)  
$i,j=1,2,3$ for scalar leptoquarks and $i,j=1,2,..8$  for scalar gluons. 
The interactions of scalar leptoquark and scalar gluon doublets with fermions in MQLS-model  
can be found in Refs.~\cite{popov-2005-20, PPSmPhAN2007}. 

\begin{figure}[htb]
\vspace*{0.5cm}
 \centerline{
\epsfxsize=1.0\textwidth
\epsffile{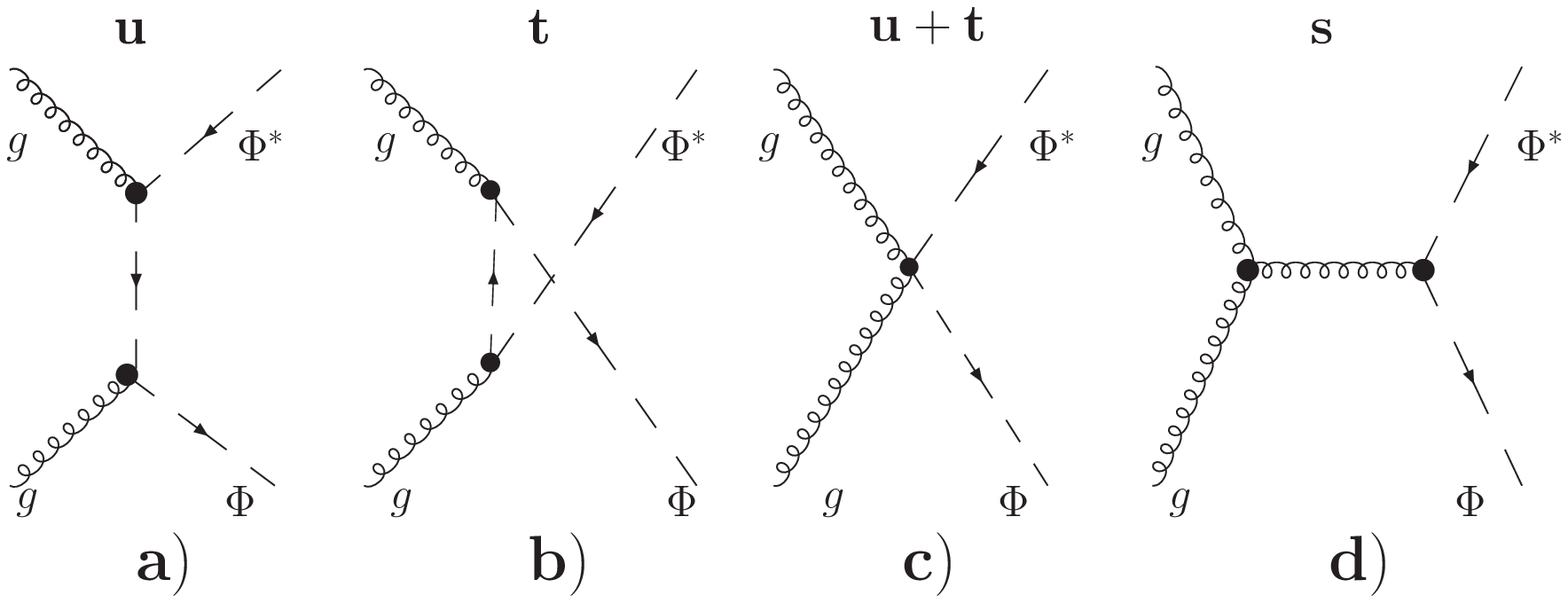} }
%\epsffile{pic1.eps} }
\vspace*{1mm}
\caption{}
\label{gg_phiphi}
%\nonumber
\end{figure}

\begin{figure}[htb]
\vspace*{0.5cm}
 \centerline{
\epsfxsize=1.0\textwidth
\epsffile{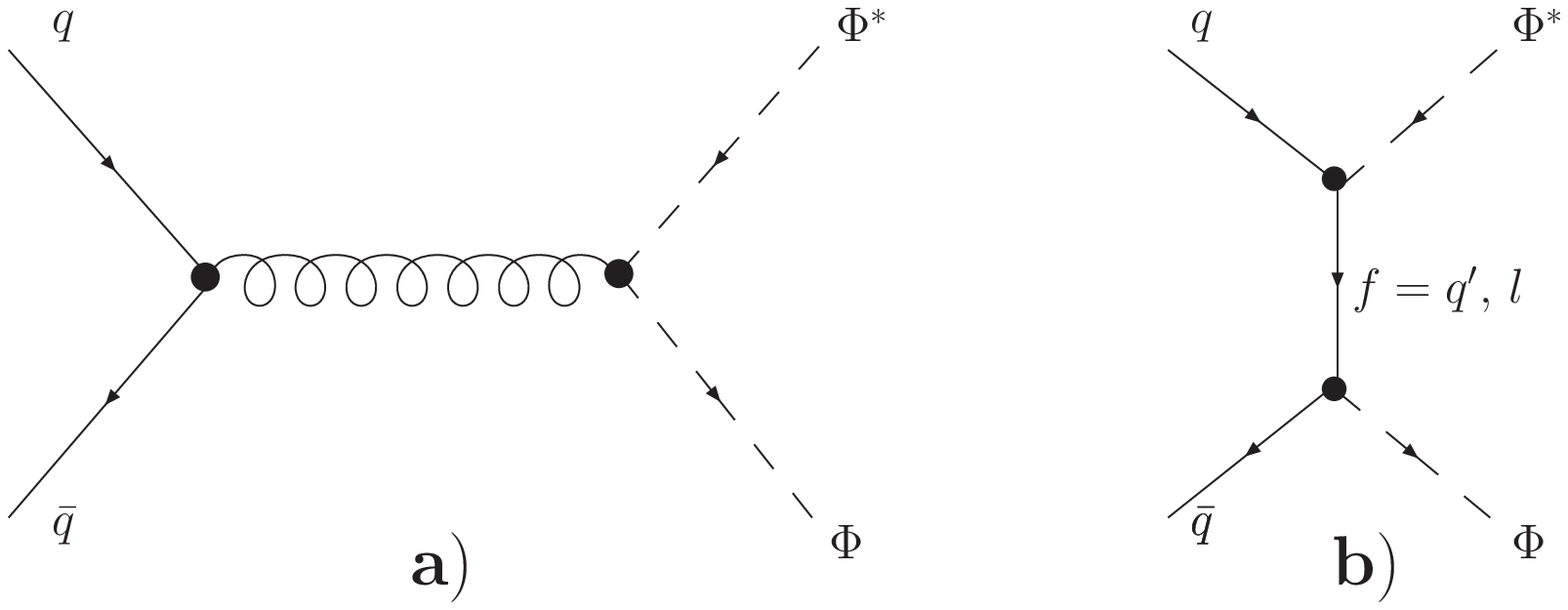}} 
%\epsffile{pic2.eps} }
\vspace*{1mm}
\caption{}
\label{qq_phiphi}
%\nonumber
\end{figure}

In $pp$-collisions colored particles can be pair produced through gluon 
fusion and through quark-antiquark pair annihilation. Contributions to the 
total production cross section for $\Phi\Phi^*$ pairs are given by the parton subprocess diagrams 
shown in Fig. \ref{gg_phiphi}, \ref{qq_phiphi}. Diagrams of the type 
\ref{qq_phiphi}-b give a small contribution to the cross section since the 
corresponding constants for scalar-fermion interactions with ordinary 
quarks and leptons in MQLS are small \cite{popov-2005-20, PPSmPhAN2007}.

%%%%%%%%%%%%%%%%%%%%%%%%%%%%%%%%%%%%%%%%%%%%%%%%%%%%%%%%%%%%%%%%%%%%%%%%%%%%
%\vspace{5mm}
%$\mathbf{gg\to \Phi^*\Phi}$

In nonabelian gauge theories the amplitudes of the scattering processes 
$2\to 2$ can be represented as the product of group and Lorentz factors 
\cite{Dongpei}. For the $\Phi^*\Phi$ production amplitude in the process 
of gluon fusion this gives 
\begin{equation}\label{ampl_dongpei}
 M(g^ag^b\to \Phi^*_i\Phi_j)\equiv M^{ab}_{ij}=\sum_pg^2_{st}(G_p)^{ab}_{ij}\frac{A_p}{C_p},
\end{equation}
where $C_p$ are the denominators 
 of the corresponding propagators 
($p=s,t,u$ is the channel index, 
$C_s\equiv\,\hat{s}=s$, 
$C_t\equiv\,\hat{t}=t-M^2_\Phi$, 
$C_u\equiv\,\hat{u}=u-M^2_\Phi$);
$(G_p)^{ab}_{ij}$ are the group factors
$(G_t)^{ab}_{ij}=(T^aT^b)_{ij}$, $(G_u)^{ab}_{ij}=(T^bT^a)_{ij}$,
$(G_{s})^{ab}_{ij}=if_{abc}T^c_{ij}$
($a,b=1..d_A$ are the color indices of colliding gluons);
%the color indices in $SU(N)$ are
%$i,j=1..3$ for the scalar leptoquarks being produced and $i,j=1..8$ for 
%scalar gluons; 
and $A_p$ is the Lorentzian part of the amplitude.

The diagrams 1-a and 1-b (fig.~\ref{gg_phiphi}) contribute  to the $u$- and $t$-channels 
of formula \eqref{ampl_dongpei} whereas the diagram 1-c gives the 
contribution to each of these channels due to the group factor of diagram 1-c $G_{t}+G_u$. 
Calculating the Lorentzian part of the amplitude gives
\begin{align}\notag
A_s=&-(q_1-q_2)\cdot\varepsilon_1(q_1,\sigma_1)(2p_1+p_2)\cdot\varepsilon_2(q_2,\sigma_2)+\\
    \label{T_s}
&+(q_1-q_2)\cdot\varepsilon_2(q_2,\sigma_2)(2p_2+p_1)\cdot\varepsilon_1(q_1,\sigma_1)+(\hat{u}-\hat{t})\varepsilon_1(q_1,\sigma_1)\cdot\varepsilon_2(q_2,\sigma_2),\\
\label{T_t}
A_t=&(2q_1-p_1)\cdot\varepsilon_1(q_1,\sigma_1)(p_2-2q_2)\cdot\varepsilon_2(q_2,\sigma_2)-\hat{t}\varepsilon_1(q_1,\sigma_1)\cdot\varepsilon_2(q_2,\sigma_2),\\
\label{T_u}
A_u=&(2q_1-p_2)\cdot\varepsilon_1(q_1,\sigma_1)(p_1-2q_2)\cdot\varepsilon_2(q_2,\sigma_2)-\hat{u}\varepsilon_1(q_1,\sigma_1)\cdot\varepsilon_2(q_2,\sigma_2),
\end{align}
where $q_i$ and $p_i$ are momenta of gluons and scalar particles 
respectively; $\varepsilon_i(q_i,\sigma_i)$ is the gluon polarization 
vector with momentum $q_i$ and polarization $\sigma_i$, $i=1,2$. 

By virtue of the masslessness of gluons we have the relation
\begin{equation}\label{massless}
    C_s+C_u+C_t=0,
\end{equation}
and taking into account the commutation relations of the $SU(N)$ group we have also the relation
\begin{equation} 
\label{G_factors} 
(G_t)^{ab}_{ij}-(G_u)^{ab}_{ij}=(G_{s})^{ab}_{ij}.
\end{equation}

From the explicit form of the amplitudes (\ref{T_s}--\ref{T_u}) it also follows that
\begin{equation}\label{T_factors}
    A_t-A_u=A_s.
\end{equation}

%The use of relations (\ref{massless}-\ref{T_factors}) substantially 
%simplifies the calculation of $M^2$.
With using of the relations (\ref{massless}--\ref{T_factors}) 
the squared amplitude \eqref{ampl_dongpei} averaged over polarizations and colors of the initial gluons 
and summed over colors of the final particles can be written as 
\begin{equation}
\label{M_square}
{\overline{|M|^2}}=\frac{1}{4}\frac{1}{d_A^2}\sum_{\sigma_1,\sigma_2}\sum_{a,b}\sum_{i,j}|M^{ab}_{ij}|^2=g_{st}^4\overline{|G|^2}\cdot\overline{\left|\frac{A_t}{C_t}+\frac{A_u}{C_u}\right|^2},
\end{equation}
where
\begin{equation}\label{G_{st}quare} \overline{|G|^2}=\frac{1}{d_A^2}\sum_{i,j,a,b} \left|\frac{C_u(G_t)^{ab}_{ij}+C_t(G_u)^{ab}_{ij}}{C_s}\right|^2=\frac{d_\Phi}{d_A^2}C_2(\Phi)
\left(
C_2(\Phi)-\frac{\hat{u}\hat{t}}{s^2}C_2(A)
\right),
\end{equation}
\begin{equation}\label{TT_square}\overline{\left|\frac{A_t}{C_t}+\frac{A_u}{C_u}\right|^2}=
\frac{1}{4}\sum_{\sigma_1,\sigma_2}
\left|\frac{A_t}{C_t}+\frac{A_u}{C_u}\right|^2=2\left[1-2\frac{sM_S^2}{\hat{u}\hat{t}} +2\left(\frac{sM_S^2}{\hat{u}\hat{t}}\right)^2\right],
\end{equation}
where $d_\Phi,d_A$ are the dimensions of the representation the field $\Phi$ 
and of the adjoint representation and $C_2(\Phi)$ and $C_2(A)$ are the 
corresponding eigenvalues of Casimir  operator. Summing over gluon 
polarization, we employ the light-cone axial gauge in which there are no 
ghost contributions \cite{Cheng}.

%%%%%%%%%%%%%%%%%%%%%%%%%%%%%%%%%%%%%%%%%%%%%%%%%%%%%%%%%%%%%%%%%%%%%%%%%%%%%%%%%%%%

With account of (\ref{M_square}--\ref{TT_square}) the differential and total $\Phi\Phi^*$ pair production 
cross sections in gluon fusion are 
\begin{equation}\label{diff_sect}    \frac{d\sigma_{gg\to\Phi\Phi^*}}{dt}=\frac{2\pi\alpha_s^2d_\Phi}{d_A^2\hat{s}^2}C_2(\Phi)
\left(
C_2(\Phi)-\frac{\hat{u}\hat{t}}{\hat{s}^2}C_2(A)
\right)
\left[1-2\frac{\hat{s}M_\Phi^2}{\hat{u}\hat{t}} +2\left(\frac{\hat{s}M_\Phi^2}{\hat{u}\hat{t}}\right)^2\right],
\end{equation}

\begin{align}\notag
 \sigma_{gg\to\Phi\Phi^*}=&\frac{\pi\alpha_s^2}{6\hat{s}}\frac{d_\Phi C_2(\Phi)}{d_A^2}
 \Bigl[C_2(A)\beta(3-5\beta^2)-12C_2\beta(\beta^2-2)+\\\label{sect}
 +&\ln\left|\frac{\beta+1}{\beta-1}\right|
 (6C_2(\Phi)(\beta^4-1)-3C_2(A)(\beta^2-1)^2)\Bigr],
 \end{align}
where $\beta=\sqrt{1-4M^2_\Phi/s}$ is the velocity of scalar particle in 
the center of mass frame, and $\hat{s}$ is the squared energy in the center of 
momentum frame of the partons. Expressions \eqref{diff_sect} and 
\eqref{sect} agree with the corresponding results of 
%\cite{manohar-2006-74} 
\cite{MW} 
and have the more simple form.

From formula \eqref{sect} for the $SU(3)$ group ($C_2(A)=3$, $d_A=8$) for 
$C_2(\Phi)=4/3$ and $d_\Phi=3$, we obtain the total production cross 
section of scalar leptoquarks $S$ 
\begin{align}\label{ss_sect}
 \sigma_{gg\to SS^*}=&\frac{\pi\alpha_s^2}{96s}
 \Bigl[\beta(41-31\beta^2)+\ln\left|\frac{\beta+1}{\beta-1}\right|
 (-\beta^4+18\beta^2-17)
 \Bigr],
 \end{align}
 which coincides with that of Ref. \cite{boos};
and for $C_2(\Phi)=3$, $d_\Phi=8$, we obtain the total production cross 
section of scalar gluons $F_a'$ in the form 

\begin{align}
 \sigma_{gg\to F_aF_a^*}=&\frac{3\pi\alpha_s^2}{16s}
 \Bigl[\beta(27-17\beta^2)+3\ln\left|\frac{\beta+1}{\beta-1}\right|
 (\beta^4+2\beta^2-3)
 \Bigr].
 \end{align}

%\vspace{5mm}
%$\mathbf{q\bar{q}\to \Phi^*\Phi}$

The production  of the scalar particles $\Phi$ in quark-antiquark pair 
annihilation are described by the diagram  shown in Fig.\ref{qq_phiphi}-a 
(the contribution of diagram \ref{qq_phiphi}-b from scalar leptoquark 
and scalar gluon doublets of MQLS - model are small). 
The corresponding cross sections are 

\begin{equation}\label{qq_diff_sect} 
\frac{d\sigma_{q\bar{q}\to\Phi\Phi^*}}{d\cos\theta}=\frac{C(\Phi)\pi \alpha_s^2}{9s}
\sin^2\hspace{-1mm}\theta\,\beta^3,
\end{equation}
\begin{equation}\label{qq_sect}
\sigma_{q\bar{q}\to\Phi\Phi^*}=\frac{4C(\Phi)\pi \alpha_s^2}{27s}\beta^3,
\end{equation}
here $C(\Phi)$ are the normalization constants of the generators for the 
representation $\Phi$: $Tr(T^a T^b)= C(\Phi)\delta^{ab}$, 
$C(\Phi)=3$ for scalar gluons and $C(\Phi)=1/2$ for scalar leptoquarks. 
For the scalar leptoquarks, expressions (\ref{qq_diff_sect}--\ref{qq_sect}) 
coincide with the corresponding expressions in \cite{boos}, but for scalar 
gluons they give cross sections 6 times larger (having the same masses) 
than those for scalar leptoquarks.

%\vspace{5mm}

\begin{figure}[htb]
\vspace*{0.5cm}
 \centerline{
\epsfxsize=1.0\textwidth 
\epsffile{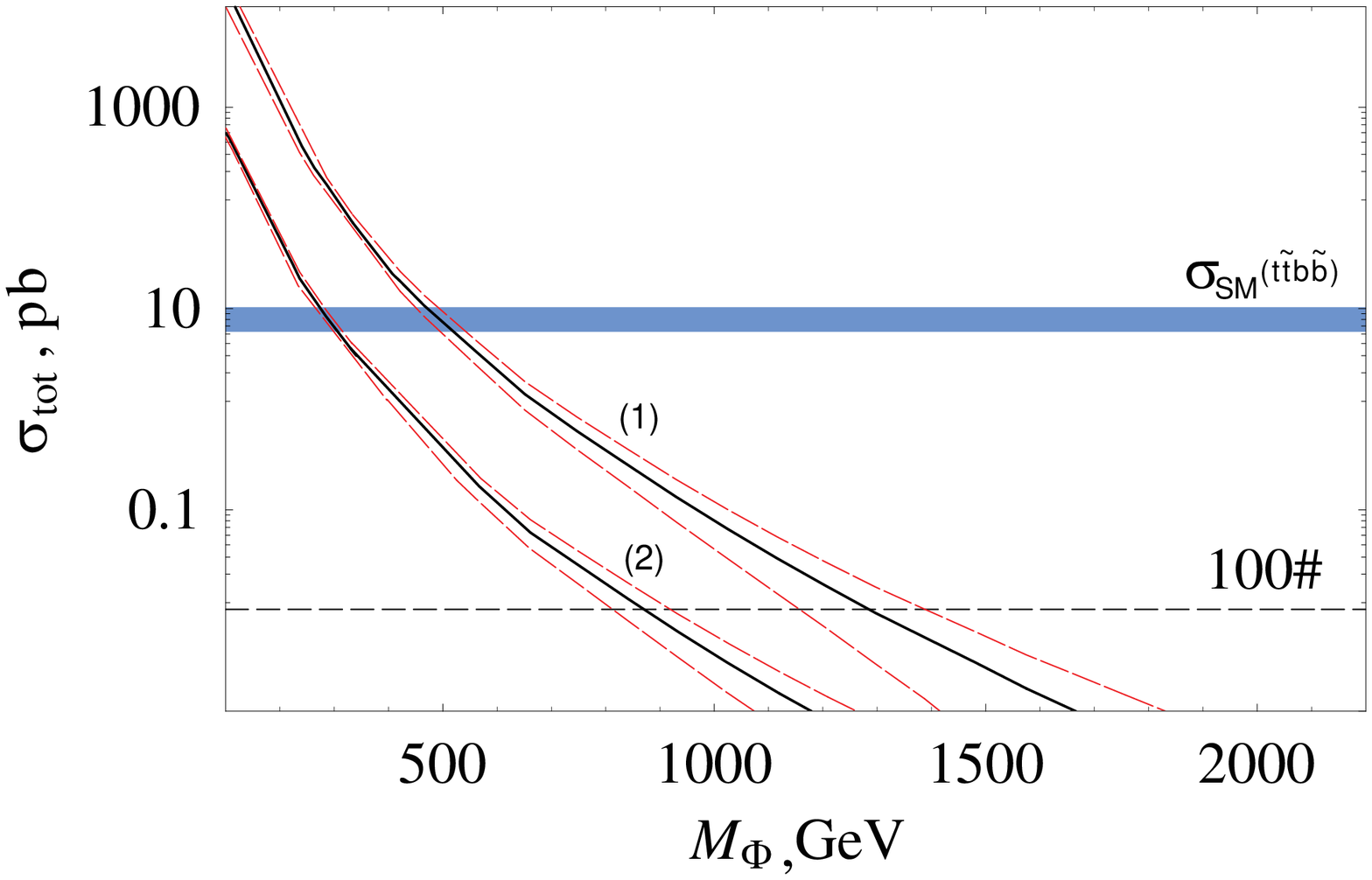}} 
%\epsffile{pic3.eps} }
\vspace*{1mm}
\caption{}
\label{ggFF_LHC}
%\nonumber
\end{figure}

The total production cross section of scalar particles $\Phi\Phi^*$ in 
$pp$-collisions at LHC energy 
%$\sigma_{tot}=\sigma(pp\to\Phi\Phi^*,\mathrm{X})$ 
$\sigma_{tot}=\sigma(pp\to\Phi\Phi^*)$ 
was calculated using the parton densities \cite{alekhin} in the leading 
order (LO) approximation with a fixed number of quark flavors. 
The total cross sections of scalar leptoquark production 
$\sigma_{tot}=\sigma(pp\to SS^*)$  and that of scalar gluon production 
$\sigma_{tot}=\sigma(pp\to FF^*)$ for LHC energy as functions of scalar 
particle masses are shown in Fig. \ref{ggFF_LHC}. 
Here and below $FF^*$ denotes $F_1F_1^*$ or $F_2F_2^*$ pairs without their summing.        

%The result of this integration (the total production cross sections in 
%the cases of scalar leptoquarks and scalar gluons as a function of scalar 
%particle mass) is shown in Fig. \ref{ggFF_LHC}. 
%
In particular, we have found that for scalar leptoquark and scalar gluon masses 
\begin{align}\label{res1}
m_S&=870^{+50}_{-60} \,\mathrm{GeV},\\\label{res2}
m_{F}&=1300^{+100}_{-130} \,\mathrm{GeV}
\end{align}
the corresponding cross sections are $\sigma(pp\to SS^*)=\sigma(pp\to 
FF^*)=0.01\,\mathrm{pb} $ (the horizontal dashed line in Fig.~\ref{ggFF_LHC}), 
which gives the number of events with production of $SS^*$ or 
$FF^*$ pairs  $N_{events}=100$ for a luminosity $L=10\, \mathrm{fb}^{-1}$. 
The errors indicated in \eqref{res1} and \eqref{res2} arise from errors in 
the distribution functions \cite{alekhin}. The result \eqref{res1} for scalar 
leptoquarks $S$ agrees with the estimate from \cite{boos}. 

For the more light masses the cross sections become larger and they are 
\begin{align}\notag
\sigma(pp\to SS^*)&=(12-0.01)\,\mathrm{pb},\\\label{res12}
\sigma(pp\to FF^*)&=(1000-0.01) \,\mathrm{pb}
\end{align}
for $m_S=250-870$ GeV and for $m_{F_a}=250-1300$ GeV respectively.

Among all the possible fermionic decays of the scalar gluons 
$F_1, \, F_{2}=(\varphi_{1}+i \varphi_{2})/\sqrt{2}$ 
of MQLS-model, the most probable are the decays 
\begin{eqnarray}
F_1\to t\tilde b,\,\,\, F_2\to t\tilde t \,\,\, (\varphi_{1}, \, \varphi_{2} \to t\tilde t) 
\label{Ftbtt}
\end{eqnarray}
with the production of the third-generation quarks 
 \cite{popov-2005-20, PPSmPhAN2007}. In the case where mass 
splitting inside the scalar doublets $\Delta m$ is small enough ($\Delta m < m_W$), 
the modes \eqref{Ftbtt} are dominant with widths of order of  
ten GeVs and with corresponding branching ratios close to unity 
$Br(F_1\to t\tilde b)  \approx Br(F_2\to t\tilde t)\approx~1$ \cite{popov-2005-20, PPSmPhAN2007}. 
If 
$\Delta m > m_W$, then the weak decays  
$$F\to F'W$$
are also possible with widths comparable to those of the decays \eqref{Ftbtt}.

Thus, observations of scalar gluons will be possible through the dominant 
events $t\tilde{t}b\tilde{b}$, $t\tilde{t}t\tilde{t}$, 
$WWt\tilde{t}b\tilde{b}$ (or
$WWt\tilde{t}t\tilde{t}$). 
For comparison with \eqref{res12} note that the SM background, for 
example,  to the  $t\tilde{t}b\tilde{b}$ events is of about 
$\sigma_{SM}(t\tilde{t}b\tilde{b})\approx 8$ pb \cite{acerMC} 
(the darkened stripe in Fig.~\ref{ggFF_LHC}). 
So, the use of the appropriate cuts can give the possibility to 
detect the events arising from the decays of the scalar gluons with their   
masses of order \eqref{res2} or below.

In conclusion, we resume the results of the work.

The cross sections of the scalar color particle production in $pp$-collisions 
are investigated. Differential and total cross sections of the 
corresponding parton subprocesses $(gg\to\Phi\Phi^*)$ and 
$(q\bar{q}\to\Phi\Phi^*)$ are obtained. The total cross sections of 
production of scalar leptoquarks $S$ and of scalar gluons $F$ in 
$pp$-collisions at the LHC are estimated and the dominant scalar gluon decay modes are discussed. 
The production cross section of scalar gluons  with masses 
$m_F\lesssim 1300$  GeV is shown to be sufficient for the effective production 
($N_{events} \gtrsim 100$ for $L=10\, \mathrm{fb}^{-1}$) 
of these particles at the LHC.

\newpage

{\Large\bf Figure captions}

\bigskip

%%%%%%%%%%%%%%%%%%%%%%%%%%%%%%%%%%%%%%%%%%%%%%%%%%%%%%%%%%%%%%%%%%%%%%%%%%%%%%%%%%%%%%%%
% чфхё№ эрўшэрхЄё  ьюш Ёшёєэъш (ьш°р)
%%%%%%%%%%%%%%%%%%%%%%%%%%%%%%%%%%%%%%%%%%%%%%%%%%%%%%%%%%%%%%%%%%%%%%%%%%%%%%%%%%%%%%%%

\begin{quotation}

\noindent
Fig. 1. Diagrams of the colored scalar particle production in $gg$-fusion. 
\end{quotation}

\begin{quotation}

\noindent
Fig. 2. Diagrams of the colored scalar particle production in $q\bar{q}$-annihilation.
\end{quotation}

\begin{quotation}
\noindent
Fig. 3. The total cross sections of pair production of scalar gluons (1) and of scalar leptoquarks (2) 
at LHC, $\sqrt{s}=14 \, TeV$.  
%   $L=10\,\mathrm{fb}^{-1}$. 
The dashed lines (1), (2) denote errors arising from pdfs.  
The horizontal dashed line shows $\sigma_{tot} = 0.01\,\mathrm{pb}$ 
which gives $N_{events}=100$ for a luminosity $L=10\, \mathrm{fb}^{-1}$. 
The horizontal darkened stripe shows the SM background for $t\tilde{t}b\tilde{b}$ events.  
\end{quotation}

%\begin{comment}

\newpage
\vspace*{5cm}
\vfill \centerline{M.V. Martynov, A.D.~Smirnov, Modern Physics Letters A}
\centerline{Fig. 1}

\newpage
\vspace*{5cm}
\vfill \centerline{M.V. Martynov, A.D.~Smirnov, Modern Physics Letters A}
\centerline{Fig. 2}

\newpage
\vspace*{5cm}
\vfill \centerline{M.V. Martynov, A.D.~Smirnov, Modern Physics Letters A}
\centerline{Fig. 3}

%\end{comment}

\end{document}